\documentclass{llncs}
\usepackage[utf8]{inputenc}
\usepackage{xspace}
\usepackage{amsmath}
\usepackage{amssymb}
\usepackage{xspace}
\usepackage{listings}
\usepackage{wasysym}
\usepackage{graphicx}
\usepackage{algorithmic,algorithm}
\usepackage{multirow}

\usepackage{theorem}
\usepackage{enumerate}

\floatname{algorithm}{Algorithm}

\newcommand{\FAdo}{{\bf FAdo}\xspace}
\newcommand{\dfa}{DFA\xspace}

\newcommand{\dfas}{DFAs\xspace}
\newcommand{\NFA}{NFA\xspace}
\newcommand{\nfas}{NFAs\xspace}

\newcommand{\RE}{\mathsf{R}}
\newcommand{\re}{RE\xspace}
\newcommand{\res}{REs\xspace}
\newcommand{\python}{\texttt{Python}\xspace}

\newcommand{\lang}{\mathcal{L}}

\newcommand{\eqRe}{=}

\newcommand{\snf}{\mbox{$\mathrm{snf}$}\xspace}
\newcommand{\snfr}{\mbox{$\mathrm{snfr}$}\xspace}
\newcommand{\alphw}{\mbox{$\mathrm{alph}$}\xspace}
\newcommand{\rpn}{\mbox{$\mathrm{rpn}$}\xspace}
\newcommand{\stc}{\mbox{$\mathrm{sc}$}\xspace}
\newcommand{\ttc}{\mbox{$\mathrm{tc}$}\xspace}
\newcommand{\isdet}{\mbox{$\mathrm{det}$}\xspace}
\newcommand{\ishom}{\mbox{$\mathrm{hom}$}\xspace}

\newcommand{\PD}{\mathsf{PD}}

\newcommand{\nfa}{\mathcal{A}}
\newcommand{\pos}{{\mathrm{pos}}}
\newcommand{\Pos}{{\mathrm{Pos}}}
\newcommand{\pd}{{\mathrm{pd}}}
\newcommand{\Alphabet}{\Sigma}
\newcommand{\letter}{\sigma}

\newcommand{\first}{\mathsf{first}}
\newcommand{\last}{\mathsf{last}}
\newcommand{\follow}{\mathsf{follow}}

\let\epsilon=\varepsilon 
\newcommand{\ommit}[1]{\ }

\title{Small \nfas from Regular Expressions:
  Some Experimental Results\thanks{This work was partially funded by
    Funda\c{c}\~ao para a Ci\^encia e Tecnologia (FCT) and Program
    POSI, and by the project ASA (PTDC/MAT/65481/2006).}}  \author{Hugo
  Gouveia\thanks{Hugo Gouveia passed away in December, 2009. Through
    2009 he was funded by a LIACC-FCT scholarship for young undergraduated researchers.},
 Nelma Moreira, Rog\'erio Reis}
\institute{DCC-FC \ \& LIACC,  Universidade do Porto \\
  R. do Campo Alegre 1021/1055, 4169-007 Porto, Portugal}

\begin{document}
\maketitle
\begin{abstract}
  Regular expressions (\res), because of their succinctness and clear
  syntax, are the common choice to represent regular
  languages. However, efficient pattern matching or word recognition
  depend on the size of the equivalent nondeterministic finite
  automata (\NFA). We present the implementation of several algorithms
  for constructing small $\epsilon$-free \nfas from \res within the
  \FAdo system, and a comparison of regular expression measures and
  \NFA sizes based on experimental results obtained from uniform
  random generated \res. For this analysis, nonredundant \res and
  reduced \res in star normal form were considered.
 \end{abstract}
\section{Introduction}
\label{sec:intro}
Regular expressions (\res), because of their succinctness and clear
syntax, are the common choice to represent regular
languages. Equivalent deterministic finite automata (\dfa) would be
the preferred choice for pattern matching or word recognition as these
problems can be solved efficiently by \dfas. However, minimal \dfas
can be exponentially bigger than \res.  Nondeterministic finite
automata (\NFA) obtained from \res can have the number of states
linear with respect to (w.r.t) the size of the \res. Because \NFA
minimization is a PSPACE-complete problem other methods must be used
in order to obtain small \nfas~usable for practical
purposes. Conversion methods from \res to equivalent \nfas can produce
\nfas without or with transitions labelled with the empty word
($\varepsilon$-\NFA). Here we consider several constructions of small
$\epsilon$-free \nfas that were recently developed or
improved~\cite{b.g.mirkin66:_algor_for_const_base_in,antimirov96:_partial_deriv_regul_expres_finit_autom_const,champarnaud02:_canon_deriv_partial_deriv_and,hromkovic01:_trans_regul_expres_into_small,ilie03:_follow,j.-m.07:_normal_expres_and_finit_autom},
and that are related with the one of Glushkov and
McNaughton-Yamada~\cite{glushkov61,mcnaughton60}. The \NFA size can be
reduced by merging equivalent
states~\cite{reduce_nfa,l.ilie05:_reduc_size_of_nfas_by}.  Another
solution is to simplify the \res before the
conversion~\cite{ellul05:_regul_expres}. Gruber and
Gulan~\cite{gruber09:_simpl_regul_expres} showed that \res in  reduced star
normal form (\snf) achieve some conversion lower bounds. Our
experimental results corroborate that \res must be converted to
reduced \snf.  In this paper we present the implementation within the
\FAdo system~\cite{fado} of several algorithms for constructing small
$\epsilon$-free \nfas from \res, and a comparison of
regular expression measures and \NFA sizes based on experimental
results obtained from uniform random generated \res. We consider
nonredundant \res and \res  in reduced \snf in particular.

\section{Regular Expressions and Finite Automata}
\label{sec:refa}
Let $\Alphabet$ be an \emph{alphabet}
(set of \emph{letters}). A \emph{word} $w$ over
$\Alphabet$ is any finite sequence of letters. The \emph{empty word}
is denoted by $\varepsilon$.  Let $\Alphabet^\star$ be the set of all
words over $\Alphabet$. A \emph{language} over $\Alphabet$ is a subset
of $\Alphabet^\star$. The set $\RE$ of \emph{regular expressions} (\re) over
$\Alphabet$ is defined by:
  $$\begin{array}{lcr}
  \alpha&:=&\emptyset \mid \varepsilon \mid \letter \in \Sigma 
\mid (\alpha + \alpha) \mid (\alpha \cdot \alpha) 
\mid \alpha^\star, \label{eq:rt}
\end{array}$$
where the operator $\cdot$ (concatenation) is often omitted.
The   language  $\lang(\alpha)$   associated   to  $\alpha\in\RE$   is
inductively   defined    as   follows:   $\lang(\emptyset)=\emptyset$,
$\lang(\varepsilon)=\{\varepsilon\}$, $\lang(\letter)=\{\letter\}$ for
$\letter                        \in                        \Alphabet$,
$\lang((\alpha+\beta))=\lang(\alpha)\cup\lang(\beta)$,
$\lang((\alpha\cdot\beta))=\lang(\alpha)\cdot\lang(\beta)$,         and
$\lang(\alpha^\star)=\lang(\alpha)^\star$.   Two  regular  expressions
$\alpha$      and      $\beta$      are      \emph{equivalent}      if
$\lang(\alpha)=\lang(\beta)$,  and  we  write $\alpha\eqRe\beta$.  The
algebraic  structure $(\RE,+,\cdot,\emptyset,\varepsilon)$ constitutes
an idempotent semiring, and with the unary operator $\star$, a Kleene
algebra.  There are  several ways  to measure  the size  of  a regular
expression.  The \emph{size} (or \emph{ordinary length}) $|\alpha|$ of
$\alpha  \in \RE$  is the  number  of symbols  in $\alpha$,  including
parentheses (but not the operator~$\cdot$); the \emph{alphabetic size}
$|\alpha|_\Alphabet$  (or $\alphw(\alpha)$) is  its number  of letters
(multiplicities included); and the \emph{reverse polish notation size}
$\rpn(\alpha)$  is the  number of  nodes  in its  syntactic tree.  The
\emph{alphabetic       size}      is      considered       in      the
literature~\cite{ellul05:_regul_expres}  the most useful  measure, and
will be the one we  consider here for several \re measure comparisons.
Moreover all these measures are  identical up a constant factor if the
regular  expression is  reduced~\cite  [Th. 3]{ellul05:_regul_expres}.
Let   $\varepsilon(\alpha)$   be   $\varepsilon$  if   $\varepsilon\in
\lang(\alpha)$,  and  $\emptyset$   otherwise.  A  regular  expression
$\alpha$  is \emph{reduced} if  it is  normalised w.r.t  the following
equivalences (rules):
$$\begin{array}{lcr}
\begin{array}{lll}
  \varepsilon\cdot\alpha&\eqRe \alpha\cdot\varepsilon&\eqRe\alpha\\
\emptyset\cdot\alpha&\eqRe \alpha\cdot\emptyset&\eqRe\emptyset\\
\emptyset + \alpha &\eqRe \alpha + \emptyset &\eqRe \alpha\\
\end{array}
&\hspace{2cm}&
\begin{array}{lll}
\varepsilon + \alpha &\eqRe \alpha + \varepsilon &\eqRe \alpha, 
\text{ where }\varepsilon(\alpha)=\varepsilon\\
\alpha^{\star\star} &\eqRe \alpha^\star\\
\emptyset^\star &\eqRe \varepsilon^\star &\eqRe \varepsilon
\end{array}
\end{array}
$$
\noindent A \re can be transformed into an equivalent reduced \re in
linear time.% on the size of the \re.

A \emph{nondeterministic automaton} (\NFA) $\nfa$ is a quintuple
$(Q,\Alphabet,\delta,q_0,F)$, where $Q$ is a finite set of states,
$\Alphabet$ is the alphabet, $\delta \subseteq Q \times \Alphabet
\times Q$ the transition relation, $q_0$ the initial state, and $F
\subseteq Q$ the set of final states. The \emph{size} of an \NFA is
$|Q|+|\delta|$. For $q\in Q$ and $\letter\in \Alphabet$, we denote by
$\delta(q,\letter)=\{p\mid(q,\letter,p)\in \delta\}$, and we can
extend this notation to $w\in \Alphabet^\star$, and to $R\subseteq Q$.
The \emph{language} accepted by $\nfa$ is $\lang(\nfa)=\{w\in
\Alphabet^\star\mid \delta(q_0,w)\cap F\not= \emptyset\}$. Two \nfas
are \emph{equivalent}, if they accept the same language.  If two \nfas
$A$ and $B$ are isomorphic, and we write $A\simeq B$.  An \NFA is
\emph{deterministic} (\dfa) if for each pair $(q,\letter) \in Q \times
\Sigma$ there exists at most one $q'$ such that $(q,\letter,q') \in \delta$.
A \dfa is \emph{minimal} if there is no equivalent \dfa with fewer
states. Minimal \dfa are unique up to isomorphism.  Given an
equivalence relation $E$ on $Q$, for $q\in Q$ let $[q]_{E}$ be the
class of $q$ w.r.t $E$, and for $T\subseteq Q$ let
$T/_{E}=\{[q]_{E}\mid q\in T\}$.  The equivalence relation $E$ is
\emph{right invariant} w.r.t an \NFA $\nfa$ if $E\subseteq (Q\setminus
F)^2\cup F^2$ and for any $p, q\in Q$, $\letter\in \Alphabet$ if $p\;
E\; q$, then $\delta(p,\letter)/_{E} = \delta(q,\letter)/_{E}$. The quotient
automaton $\nfa/_E=(Q/_E,\Alphabet,\delta_E,[q_0]_E,F/_E)$, where
$\delta_E = \{([p]_E,\letter,[q]_E) \mid (p,\letter,q) \in \delta\}$,
satisfies $\lang(\nfa)=\lang(\nfa/_E)$. Given two equivalence
relations over a set $Q$, $G$ and $H$, we say that $G$ is \emph{finer}
than $H$ (and $H$ \emph{coarser} than $G$) if and only if $G\subseteq
H$.

\section{Small \nfas from Regular Expressions}
\label{sec:nfas}
We consider three methods for constructing small \nfas $\nfa$ from a
regular expression $\alpha$ such that $\lang(\nfa)=\lang(\alpha)$, i.e.,
they are \emph{equivalent}.

\subsection{Position Automata}
\ommit{ Given a regular expression $\alpha$, it is possible to
  recognize a word of $\lang(\alpha)$ by following the expression
  symbols in an adequate manner. For instance the word $aabab$ can be
  obtained from $a^\star(ba^\star)^\star$ by visiting two times the
  first $a$ and two times the subexpression $(ba^\star)$, visiting
  $b$, visiting the second $a$, visiting $b$ again and finally
  visiting zero times $a$.  It is obvious that the position of each
  letter is important for the word recognition.  } The position
automaton construction was independently proposed by Glushkov, and
McNaughton and Yamada~\cite{glushkov61,mcnaughton60}. Let
$\Pos(\alpha)=\{1,2,\ldots,|\alpha|_\Alphabet\}$ for $\alpha\in \RE$,
and let $\Pos_0(\alpha)=\Pos(\alpha)\cup\{0\}$. We consider the
expression $\overline{\alpha}$ obtained by marking each letter
$\letter$ with its position $i$ in $\alpha$, $\letter_i$. The same
notation is used to remove the markings, i.e.,
$\overline{\overline{\alpha}}=\alpha$.  For $\alpha\in\RE$ and $i \in
\Pos(\alpha)$, let $\first(\alpha)=\{j\mid \exists w\in
\overline{\Alphabet}^\star, \letter_jw\in \lang(\overline{\alpha})\}$,
$\last(\alpha)=\{j\mid \exists w\in \overline{\Alphabet}^\star,
w\letter_j\in \lang(\overline{\alpha})\}$, and
$\follow(\alpha,i)=\{j\mid \exists u,v\in \overline{\Alphabet}^\star,
u\letter_i\letter_jv\in \lang(\overline{\alpha})\}$. Let
$\follow(\alpha,0)=\first(\alpha)$. The \emph{position automaton} for
$\alpha\in\RE$ is
$\nfa_\pos(\alpha)=(\Pos_0(\alpha),\Alphabet,\delta_\pos,0,F)$, with
$\delta_\pos=\{(i,\overline{\letter_j},j)\mid\;
j\in\follow(\alpha,i)\}$ and $F=\last(\alpha)\cup\{0\}$ if
$\varepsilon(\alpha)=\varepsilon$, and $F=\last(\alpha)$,
otherwise. We note that the number of states of $\nfa_\pos(\alpha)$ is
exactly $|\alpha|_\Alphabet+1$. Other interesting property is that
$\nfa_\pos$ is \emph{homogeneous}, i.e., all transitions arriving at a
given state are labelled by the same
letter. Brüggemann-Klein~\cite{bruggemann-klein93:_regul_expres_into_finit_autom}
showed that the construction of $\nfa_\pos$ can be obtained in ${\cal
  O}(n^2)$ ($n=|\alpha|$)
%if it is ensured that in the computation of
%the $\first$, $\last$ and $\follow$ sets, all the unions are
%disjoint. That can be achieved 
if the regular expression $\alpha$ is
in the so called \emph{star normal form} (\snf), i.e., if for each
subexpression $\beta^\star$ of $\alpha$, $\forall x\in
\last(\beta),\,\follow(\beta,x)\cap \first(\beta)=\emptyset$ and
$\varepsilon(\beta)=\emptyset$. For every $\alpha\in \RE$ there is an
equivalent \re in star normal form $\alpha^\bullet$ that can be
computed in linear time and such that
$\nfa_\pos(\alpha)\simeq\nfa_\pos(\alpha^\bullet)$.%~\cite[Thm.~3.1]{bruggemann-klein93:_regul_expres_into_finit_autom}.

\subsection{Follow Automata}
\label{sec:f}
Ilie and Yu~\cite{ilie03:_follow} introduced the construction of the
follow automaton from a \re. Their initial algorithm begins by
converting $\alpha\in\RE$ into an equivalent $\varepsilon$-\NFA from
which the follow automaton $\nfa_f(\alpha)$ is obtained. For
efficiency reasons we implemented that method in the \FAdo
library. The \emph{follow automaton} is a quotient of the position
automaton w.r.t the right-invariant equivalence given by the
\emph{follow relation} $\equiv_f\subseteq\Pos_0^2$ that is defined by:
\begin{center}
\begin{tabular}{lcr}
$\forall x,y \in \Pos_0(\alpha), x\equiv_f y$& if&
\begin{tabular}[t]{l}
  (i) both $x,y$ or none belong to $\last(\alpha)$ and\\
(ii) $\follow(\alpha,x)=\follow(\alpha,y)$ 
\end{tabular}
\end{tabular}
 \end{center}
\begin{proposition}[Ilie and Yu, Thm.~23] $\nfa_f(\alpha)\simeq
  \nfa_\pos(\alpha)/_{\equiv_f}$.
\end{proposition}

\subsection{Partial Derivative Automata}
\label{sec:pd}
Let $S \cup \{\beta\}$ be a set of regular expressions. Then $S \odot
\beta = \{ \alpha\beta \, | \, \alpha \in S \}$ if $\beta \not=
\emptyset$ and $S \odot \emptyset = \emptyset$.  For $\alpha\in\RE$
and $\letter \in \Alphabet $, the set $\partial_\letter(\alpha)$ of
\emph{partial derivatives} of $\alpha$ w.r.t. $\letter$ is defined
inductively as follows:
{\small $$\begin{array}{ll}
\begin{array}{ll}
  \partial_\letter(\emptyset) &= \partial_\letter(\varepsilon) =
  \emptyset\\
  \partial_\letter(\letter') &= \left\{
  \begin{array}{ll}
    \{\varepsilon\} & \text{if } \letter' \equiv \letter\\
    \emptyset & \text{otherwise}
  \end{array}
  \right.\\
  \partial_\letter(\alpha^\star) &= \partial_\letter(\alpha)\odot\alpha^\star
\end{array} \ \ \ \ \ \ \ \ 
\begin{array}{ll}
  \partial_\letter(\alpha+\beta) &= \partial_\letter(\alpha) \cup
  \partial_\letter(\beta)\\
  \partial_\letter(\alpha\beta) &= \left\{  \begin{array}{ll}
     \partial_\letter(\alpha)\odot\beta \cup
 \partial_\letter(\beta) & \text{if }  \varepsilon(\alpha)=  \varepsilon\\
    \partial_\letter(\alpha)\odot\beta & \text{otherwise.}
 \end{array} \right.
\end{array}
\end{array}$$}

\noindent This definition can be extended to sets of regular
expressions, words, and languages.  Given $\alpha\in\RE$ and
$\letter\in \Alphabet$, $\partial_\letter(S) = \cup_{\alpha\in
  S}\partial_\letter(\alpha)$ for $S\subseteq \RE$,
$\partial_\varepsilon(\alpha) = \{\alpha\}$,
$\partial_{w\letter}(\alpha) = \partial_\letter(\partial_w(\alpha))$
for $w\in \Alphabet^\star$, and $\partial_{L}(\alpha) = \cup_{w\in
  L}\partial_w(\alpha)$ for $L\subseteq\Alphabet^\star$.  The
\emph{set of partial derivatives} of $\alpha$ is denoted by
$\PD(\alpha)=\{\partial_{w}(\alpha)\mid w\in \Alphabet^\star\}.$

Given a regular expression $\alpha$, the partial derivative automaton
$\nfa_\pd(\alpha)$, introduced by Mirkin and
Antimirov~\cite{b.g.mirkin66:_algor_for_const_base_in,antimirov96:_partial_deriv_regul_expres_finit_autom_const},
is defined by
\begin{center}$
\nfa_\pd(\alpha) = (\PD(\alpha),\Alphabet ,\delta_{pd},\alpha,\{q \in \PD(\alpha)\mid \varepsilon(q)=\varepsilon\}),$
\end{center}
\noindent where $\delta_{pd}(q,\sigma)=\partial_\sigma(q)$, for all
$q\in\PD(\alpha)$ and $\letter \in \Alphabet$.
\begin{proposition}[Mirkin and Antimirov] $\lang(\nfa_\pd(\alpha))=\lang(\alpha).$
\end{proposition}
\noindent Champarnaud and
Ziadi~\cite{champarnaud02:_canon_deriv_partial_deriv_and} showed that
the partial derivative automaton is also a quotient of the position
automaton. Champarnaud \emph{et
  al.}~\cite{j.-m.07:_normal_expres_and_finit_autom} proved that for
\re reduced and in star normal form  the size of
its partial derivative automaton $\nfa_\pd$ is always smaller than the one of
its follow automaton $\nfa_f$.
\subsection{Complexity}
\label{ref:complexity}
The automata here presented $\nfa_\pos$, $\nfa_f$ and $\nfa_\pd$ can
in worst-case be constructed in time and space ${\cal O}(n^2)$, and
have, in worst-case, size ${\cal O}(n^2)$, where $n$ is the size of
the \re. Recently,
Nicaud~\cite{nicaud09:_averag_size_of_glush_autom_c} showed that on
the average-case the size of the $\nfa_\pos$ automata is linear.
The best worst case construction of
$\epsilon$-free \nfas from \re is the one presented by Hromkovic
\emph{et al.}~\cite{hromkovic01:_trans_regul_expres_into_small} that
can be constructed and have~size ${\cal O}(n(\log n ^2))$. However
this construction is not considered here.

\section{\nfas Reduction with Equivalences}

It is possible to obtain in time ${\cal O}(n\log n)$ a (unique)
minimal \dfa equivalent to a given one. However \NFA state minimization  is
PSPACE-complete and, in general, minimal \nfas are not unique.
Considering the exponential succinctness of \nfas w.r.t \dfas, it is
important to have methods to obtain small \nfas.  Any right-invariant
equivalence relation over $Q$ w.r.t $\nfa$ can be used to diminish
the size of $\nfa$ (by computing the quotient automaton). The
\emph{coarsest right-invariant equivalence} $\equiv_R$ can be computed
by an algorithm similar to the one used to minimize
\dfas~\cite{reduce_nfa}. This coincides with the notion of
(auto)-bisimulation, widely applied to transition systems and which
can be computed efficiently (in almost linear time) by the Paige and Tarjan
algorithm~\cite{tarjan87}. A \emph{left-invariant} equivalence relation on $Q$
w.r.t $\nfa$ is any right-invariant equivalence relation on the
reversed automaton of $\nfa$,
$\nfa^r=(Q,\Alphabet,\delta_r,F,\{q_0\})$, where $q\in
\delta^r(p,\letter)$ if $p\in \delta(q,\letter)$ (and we allow
multiple initial states). The \emph{coarsest left-invariant
  equivalence} on $Q$ w.r.t $\nfa$, $\equiv_L$, is  $\equiv_R$~of~$\nfa^r$.
\section{\FAdo Implementations}
\label{sec:fado}
\begin{figure}
\vspace{-0.2cm}
\begin{center}
\includegraphics[width=7cm]{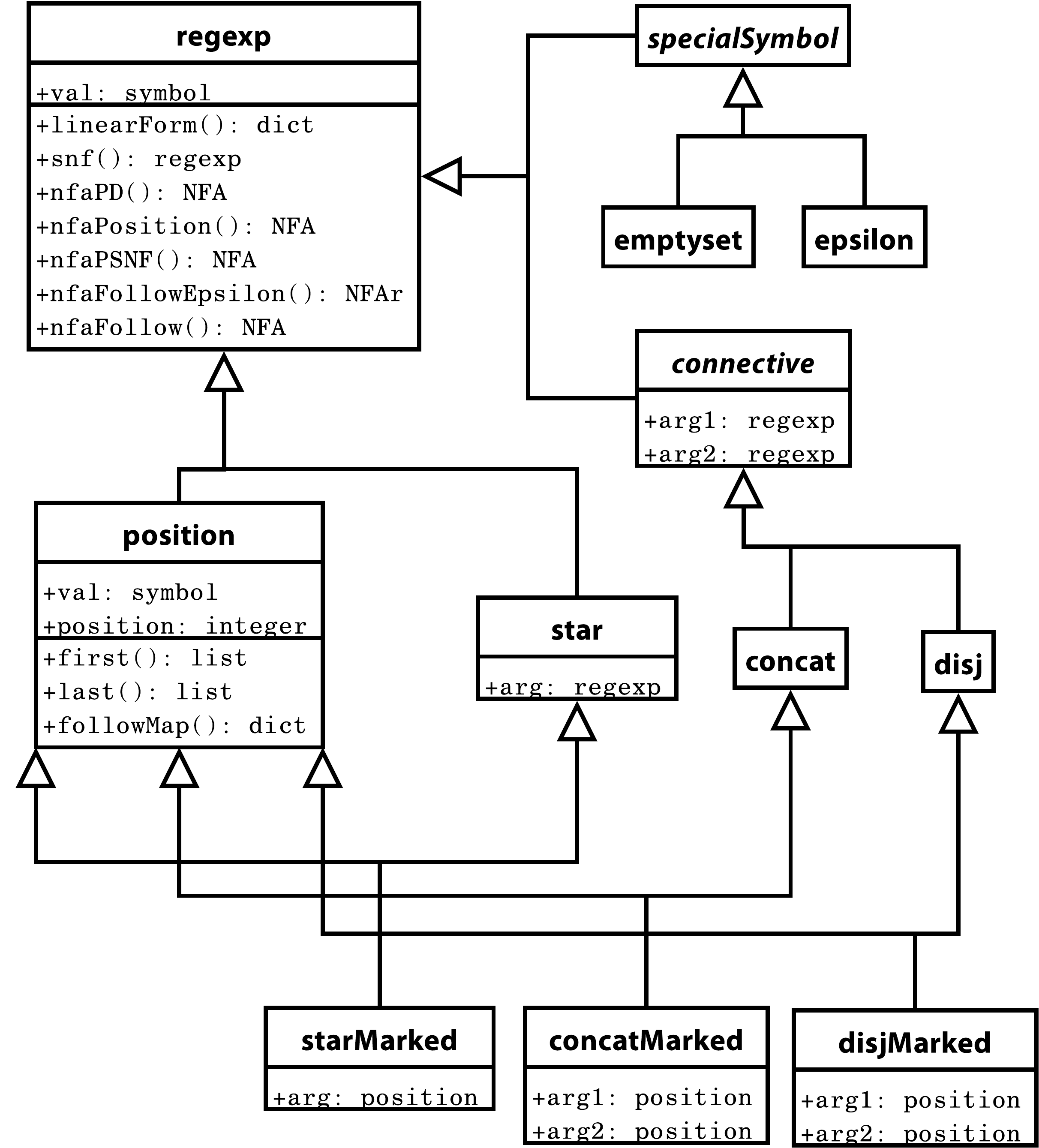}   
\end{center}
\vspace{-0.2cm}
\caption{\FAdo classes for \res}
\label{fig:regexpclass}
\vspace{-0.2cm}
\end{figure}

\FAdo~\cite{fado,moreira05_c:_inter_manip_regul_objec_fado,almeida09:_fado_guitar_c}
is an ongoing project that aims to provide a set of tools for symbolic
manipulation of formal languages. To allow high-level programming with
complex data structures, easy prototyping of algorithms, and
portability are its main features. It is mainly developed in the
\python programming language.  In \FAdo, regular expressions and
finite automata are implemented as \python classes.

Figure~\ref{fig:regexpclass} presents the classes for \res and the
main methods described in this paper.  The \texttt{regexp} class is
the base class for all \res and the class \texttt{position} is the
base class for marked \res. The methods \texttt{first()},
\texttt{last()} and \texttt{followMap()} (where
$\follow(\alpha,x)=\{\beta\mid (x,\beta)\in \texttt{followMap()}\}$)
are coded for each subclass. The method \texttt{nfaPosition()}
implements a construction of the $\nfa_\pos$ automaton without
reduction to \snf. Brüggemann-Klein algorithm is implemented by the
\texttt{nfaPSNF()} method.  The methods \texttt{nfaFollowEpsilon()}
and \texttt{nfaFollow()} implement the construction of the $\nfa_f$ via
an $\epsilon$-\NFA{}. The exact text of all these  algorithms is too long to present
here.  The method \texttt{nfaPD()} computes the $\nfa_\pd$ and uses
the method \texttt{linearForm()}. This method  implements the function
$\mathsf{lf}()$ defined by
Antimirov~\cite{antimirov96:_partial_deriv_regul_expres_finit_autom_const}
to compute the partial derivatives of a \re w.r.t all letters. 
Algorithm~\ref{alg:nfapd} presents the computation of the $\nfa_\pd$.

{\small
\vspace{-0.5cm}
\begin{algorithm}[H]
  \begin{algorithmic}[]
    \STATE $Q \leftarrow \{\alpha\}$
    \STATE $\delta \leftarrow \emptyset$
    \STATE $F \leftarrow \emptyset$
    \STATE stack $\leftarrow \{\alpha\}$
    \WHILE{pd $\leftarrow$ POP(stack)}
    \FOR{(head,tail) $\in \mathsf{lf}($pd$)$}
    \IF{$\neg$ tail $\in Q$}
    \STATE $Q \leftarrow Q \cup \{$tail$\}$
    \STATE PUSH(stack,pd)
    \ENDIF
    \STATE $\delta($pd,head$) \leftarrow \delta($pd,head$) \cup \{$tail$\}$
    \ENDFOR
    \IF{$\epsilon($pd$)$}
    \STATE $F \leftarrow F \cup \{$pd$\}$
    \ENDIF
    \ENDWHILE
  \end{algorithmic}
  \caption{Computation of $\nfa_\pd$}
  \label{alg:nfapd}
\end{algorithm}
}
\vspace{-0.5cm}

\begin{figure}
\begin{center}
  \includegraphics[width=7cm]{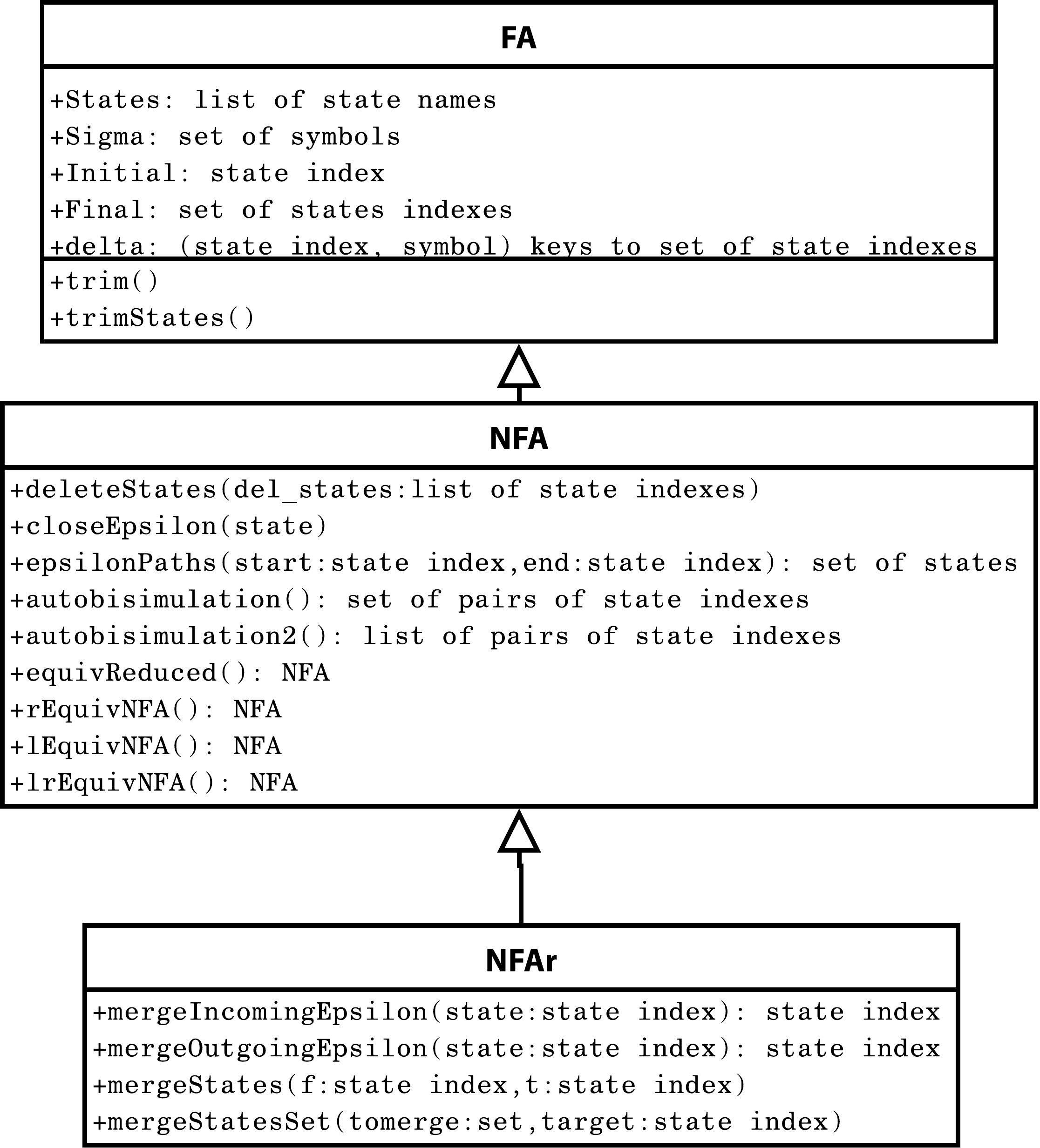}
\end{center}
\caption{\FAdo classes for \nfas}
\label{fig:faclass}
\vspace{-1cm}
\end{figure}

Figure~\ref{fig:faclass} presents the classes for finite automata.
\texttt{FA} is the abstract class for finite automata. The class
\texttt{NFAr} includes the inverse of the transition relation, that is
not included in the \texttt{NFA} class for efficiency reasons. In the
\texttt{NFA} class the method \texttt{autobisimulation()} implements a
na\"ive version for compute $\equiv_R$, as presented in
Algorithm~\ref{alg:equiv}. Given an equivalence
relation the method \texttt{equivReduced()} builds the quotient
automaton.  Given an \NFA \textbf{A}, \textbf{A}.\texttt{rEquiv()}
corresponds to $\nfa/_{\equiv_R}$, \textbf{A}.\texttt{lEquiv()} to
$\nfa/_{\equiv_L}$ and \textbf{A}.\texttt{lrEquiv()} to
$(\nfa/_{\equiv_L})/_{\equiv_R}$. We refer the reader to
Gouveia~\cite{gouveia09:_de_initos_pequen} and to \FAdo
webpage~\cite{fado} for more implementation details.

{\small \vspace{-0.3cm}\begin{algorithm}[H]
 \begin{algorithmic}[]
    \STATE $\overline{R} \leftarrow \emptyset$
    \FOR{$(p,q) \in Q \times Q$}
    \IF{$p \in F \not\Leftrightarrow q \in F$}
    \STATE $\overline{R} \leftarrow \overline{R} \cup \{(p,q)\}$
    \ENDIF
    \ENDFOR
    \WHILE{$\exists (x,y) \not\in \overline{R} \colon \exists{\letter\in\Alphabet}\colon\exists{
      z\in\delta(x,\letter)}\colon\forall{w \in\delta(y, \letter)}\colon z
      \overline{R} w$}
    \STATE $\overline{R} \leftarrow \overline{R} \cup \{(x,y),(y,x)\}$
    \ENDWHILE
    \STATE $R \leftarrow (Q\times Q)\setminus \overline{R}$
    \STATE \textbf{Return} $R$
  \end{algorithmic}
  \caption{Computation of the set $R$ corresponding to $\equiv_R$.}% for  $\nfa=(Q,\Alphabet,\delta,q_0,F)$}
  \label{alg:equiv}
\end{algorithm}}

\section{\re Random Generator}
\label{sec:radom}
Uniform random generators are essential to obtain reliable
experimental results that can provide information about the
average-case analysis of both computational and descriptional
complexity.  For general regular expressions, the task is somehow
simplified because they can be described by small unambiguous
context-free grammars from which it is possible to build uniform
random
generators~\cite{mairson94:_gener}. %,p.94:_calcul_for_random_gener_of}.
%Unfortunately, general \res are highly redundant, as equivalent \res
%can be syntacticly completely different
%~\cite{ellul05:_regul_expres,lee05:_enumer_regul_expres_and_their_languag_c}.
In the \FAdo system we implemented the method described by Mairson
\cite{mairson94:_gener} for the generation of context-free
languages. The method accepts as input a context-free grammar and the
size of the words to be uniformly random generated.

The random samples need to be consistent and large enough to ensure
statistically significant results. To have these samples readily
available, the \FAdo system includes a dataset of random \re, that can
be accessed online. The current dataset was obtained using a grammar
for \res given by Lee and Shallit
~\cite{lee05:_enumer_regul_expres_and_their_languag_c}, and that is
presented in Figure~\ref{fig:grammar}. This grammar generates \res
normalized by rules that define reduced \res, except for certain cases
of the rule: $\varepsilon+\alpha$, where
$\varepsilon(\alpha)=\varepsilon$. The database makes available random
samples of \res with different sizes between $10$ and $500$ and with
alphabet sizes between $2$ and~$50$.
\begin{figure}
  \centering
  {\small$
  \begin{array}{lcl}
    S &:=& A \mid C \mid E \mid \Alphabet \mid \varepsilon \mid \emptyset\\
                          C &:=& C\; R \mid R\; R\\
                          R &:=& (\; A\; ) \mid E \mid \Alphabet \\
                          E &:=& (\; A \;)^\star \mid (\; C\; )^\star \mid \Alphabet ^\star\\
                          A &:=& \varepsilon\; +\; X \mid Y + Z\\
                          X &:=& T \mid T\; +\; X\\
                          T &:=& C \mid \Alphabet\\
                          Y &:=& Z \mid Y\; +\; Z\\
                          Z &:=& C \mid E \mid \Alphabet
  \end{array}
$}
  \caption{Grammar for almost reduced \res. The start symbol is $S$.}
  \label{fig:grammar}
\end{figure}

\section{Experimental Results}

In order to experiment with several properties of \res and \nfas we
developed a generic program to ease to add/remove the methods to be
applied and to specify the data, from the database, to be used. Here
we are interested in the comparison of several \res descriptional
measures with measures of the \nfas obtained using the methods earlier
described.

For \res we considered the following properties: the alphabetic size
(\alphw); the \rpn size (\rpn); test if it is in \snf (\snf); if not in
\snf, compute the \snf and its measures (\alphw,\rpn); test if it is
reduced; if not reduced, reduce it and compute its measures
(\alphw,\rpn); the number of states (\stc) and number of transitions
(\ttc) of the equivalent minimal \dfa. 

For each \NFA ($\nfa_\pos$,
$\nfa_f$, and $\nfa_\pd$) we considered the following properties: the
number of states ($|Q|$); the number of transitions ($|\delta|$); if
it is deterministic (\isdet); and if it is homogeneous (\ishom). All
these properties were also considered for the case where the \res are
in \snf, and for the \nfas obtained after applying the invariant
%equivalences $\equiv_R$ (\riequiv), $\equiv_L$ (\liequiv), and both of
%them (\lrequiv).
equivalences $\equiv_R$, $\equiv_L$, and their composition.

All tests were performed on samples of $10,000$ uniformly random
generated \res. Each sample contains \res of size $50$, $100$, $200$
and $300$, respectively.

Table~\ref{tab:resp} shows some results concerning \res.  The ratio of
alphabetic size to \rpn size is almost constant for all
samples. Almost all \res are in \snf, so we do not
presented the measures after transforming into \snf. This fact is
relevant as the \res were generated only \emph{almost reduced}. The
column \snfr contains the percentage of \res for which their \snf are
reduced. It is interesting to note that the average number of states
of the minimal \dfa (\stc) is near \alphw{} (i.e., near the number of
states of $\nfa_\pos$). The standard deviation is here very high. For
the sample of size $300$, however, $99\%$ of the \res have $160\leq
\stc \leq 300$. More theoretical work is needed for a deeper
understanding of these results.

\begin{table}
\vspace{-0.3cm}
  \caption{Statistical values for \re measures, where (avg) is the average and (std) the standard deviation.}
  \label{tab:resp}
  \centering
{\small
  \begin{tabular}{|r|||r|r||r|r||r||r|r||r|r||r|r||r|r|}\hline{}\multirow{2}{5mm}{size}&\multicolumn{2}{c||}{\alphw}&\multicolumn{2}{c||}{\rpn}&\multirow{2}{7mm}{$\frac{\rpn}{\alphw}$}&\multirow{2}{6mm}{\snf}&\multirow{2}{7mm}{\snfr}&\multicolumn{2}{c||}{\stc}&
  \multicolumn{2}{c||}{\ttc}&
  \multirow{2}{7mm}{$\frac{\stc}{\alphw}$}&\multirow{2}{7mm}{$\frac{\ttc}{\alphw}$}\\\cline{2-5}\cline{9-12}
  &avg&std&avg&std&&&&avg&std&avg&std&&\\\hline\hline
  $50$  &$42$  &$6.39$  &$85$   &$10.80$  &$2.04$  &$97\%$  &$99\%$&$38$&$9.42$&$44$&$6.39$&$0.92$&$1.05$\\
  $100$ &$77$  &$10.26$ &$161$  &$17.41$  &$2.08$  &$93\%$  &$98\%$&$69$&$20.00$&$89$&$37.47$&$0.89$&$1.15$\\
  $200$ &$165$ &$25.75$ &$340$  &$43.83$  &$2.06$  &$90\%$  &$97\%$&$160$&$91.58$&$203$&$186.10$&$0.97$&$1.24$\\
  $300$ &$247$ &$38.06$ &$511$  &$64.96$  &$2.06$  &$87\%$  &$95\%$&$258$&$300.01$&$343$&$617.51$&$1.04$&$1.4$\\
  \hline
\end{tabular}}
\vspace{-0.5cm}
\end{table}

Table~\ref{tab:nfasp} and Table~\ref{tab:ratios} show some results
concerning the \nfas obtained from \res.  In Table~\ref{tab:nfasp} the
values not in percentage are average values.  If $\nfa_\pos$ is
deterministic then the \res is unambiguous (and strong unambiguous, if
in \snf)~\cite{bruggemann-klein93:_regul_expres_into_finit_autom}. The
results obtained suggest that perhaps $25\%$ of the reduced \res are
strong unambiguous. Note that if $\nfa_\pos$ is not deterministic,
almost certainly, neither $\nfa_\pd$ nor $\nfa_f$ are. For reasonable
sized \res, although $\nfa_\pos$ are homogeneous it is unlikely that
either $\nfa_\pd$ or $\nfa_f$ will be so.  It is not significant the
difference between $|Q_f|$ and $|Q_\pd|$. On
average $|\delta_\pos|$ seems linear in the size of the \re, and that
fact was recently proved by
Nicaud~\cite{nicaud09:_averag_size_of_glush_autom_c}.

\begin{table}
\vspace{-0.5cm}
\centering  {\small
  \caption{\NFA measures.}
  \label{tab:nfasp}
  \begin{tabular}{|r|||r|r|r|r||r|r|r|r||r|r|r|r|}
    \hline\multirow{2}{5mm}{size}&\multicolumn{4}{c||}{$\nfa_\pos$}&\multicolumn{4}{c||}{$\nfa_f$}&\multicolumn{4}{c|}{$\nfa_\pd$}\\\cline{2-13}
    &$|Q_\pos|$&$|\delta_\pos|$&\isdet&\ishom&$|Q_f|$&$|\delta_f|$&\isdet&\ishom&$|Q_\pd|$&$|\delta_\pd|$&\isdet&\ishom\\\hline
$50$&$43$&$51$&$49.1\%$&$100\%$&$38$&$44$&$49.3\%$&$13.7\%$&$38$&$44$&$49.4\%$&$13.6\%$\\
$100$&$78$&$104$&$16.0\%$&$100\%$&$67$&$84$&$17.0\%$&$1.0\%$&$66$&$83$&$17.0\%$&$1.0\%$\\
$200$&$166$&$211$&$27.6\%$&$100\%$&$148$&$175$&$27.7\%$&$1.5\%$&$146$&$173$&$27.7\%$&$1.4\%$\\
$300$&$248$&$317$&$23.9\%$&$100\%$&$222$&$262$&$23.9\%$&$0.5\%$&$220$&$260$&$23.9\%$&$0.5\%$\\\hline
  \end{tabular}
}
\vspace{-1cm}
\end{table}

\begin{table}
  \caption{Ratios of \NFA measures.}
  \label{tab:ratios}
  \centering
{\small
  \begin{tabular}{|r||r|r|r|r|r|r|r|r|}
    \hline{}
    size&$\frac{|\delta_\pos|}{\alphw+1}$&$\frac{|Q_f|}{/\alphw+1}$&$\frac{|\delta_f|}{\alphw+1}$ &$\frac{|Q_\pd|}{\alphw+1}$&$\frac{|\delta_\pd|}{\alphw+1}$&$\frac{|\delta_\pd|}{|\delta_\pos|}$&$\frac{|Q_\pd|}{|Q_f|}$&$\frac{|\delta_\pd|}{|\delta_f|}$\\\hline
    $50$&$1.18$&$0.90$&$1.02$&$0.89$&$1.02$&$0.86$&$0.99$&$0.99$\\
    $100$&$1.33$&$0.85$&$1.07$&$0.84$&$1.05$&$0.79$&$0.98$&$0.99$\\
    $200$&$1.27$&$0.89$&$1.06$&$0.88$&$1.05$&$0.82$&$0.99$&$0.99$\\
    $300$&$1.28$&$0.89$&$1.06$&$0.88$&$1.05$&$0.82$&$0.99$&$0.99$\\\hline
  \end{tabular}
}
\vspace{-0.5cm}
\end{table}
Reductions by $\equiv_R$ and $\equiv_L$ (or
$\equiv_R\circ \equiv_L$) decrease by less than $2\%$ the size of the
considered \nfas ($\nfa_\pos$, $\nfa_f$, and $\nfa_\pd$). In
particular the quotient automata of $\nfa_\pos$ are less than $1\%$
smaller than $\nfa_\pd$.  In general, we can hypothesize that
reductions by the coarsest invariant equivalences are not significant
when \res are reduced (and/or are in \snf).

\section{Conclusion}

We presented a set of tools within the \FAdo system to uniformly random
generate \res, to convert \res into $\epsilon$-free \nfas and to 
simplify both \res and \nfas.  These tools can be
used to obtain experimental results about the relative descriptional
complexity of regular language representations on the average
case. Our experimental data corroborate some previous experimental and
theoretical results, and suggest some new hypotheses to be
theoretically proved. We highlight the two following
conjectures. Reduced \res have high probability of being
in \snf. And the $\nfa_\pd$ obtained from \res in reduced \snf seems to almost coincide
with quotient automata of $\nfa_\pos$ by $\equiv_R\circ \equiv_L$.
\medskip

We would like to thank the anonymous referees for their
comments that helped to improve this paper.  
\bibliographystyle{alpha}
\newcommand{\etalchar}[1]{$^{#1}$}

%\bibliography{FAdoBib}
\end{document}